\documentstyle[epsfig]{aipproc}

\begin{document}

\begin{flushright}
{\small
SLAC--PUB--8745\\
December 2000\\}
\end{flushright}

\title{Experimental Issues for Precision Electroweak Physics at a High-Luminosity Z Factory}

\author{P. C. Rowson$^*$ and M. Woods}
\address{Stanford Linear Accelerator Center, Stanford, CA 94309\\
         $^*$conference speaker }

\maketitle

\newcommand{\ee}{\mbox{$e^+e^-$}}
\newcommand{\phinaught}{\mbox{$\langle\Phi\rangle_0$}}
\newcommand{\mt}{m_t}
\newcommand{\mh}{m_H}
\newcommand{\mgev}{GeV/c$^2$}
\newcommand{\mevocsq}{MeV/c$^2$}
\newcommand{\mum}{$\mu$m}
\newcommand{\roots}{\sqrt{s}}
\newcommand{\ALR}{A_{LR}}
\newcommand{\alro}{A_{LR}^0}
\newcommand{\swein}{\sin^2\theta_W^{\rm eff}}
\newcommand{\pole}{{\cal P}_e}
\newcommand{\polp}{{\cal P}_p}
\newcommand{\pollin}{{\cal P}_e^{lin}}
\newcommand{\pollum}{\bar{\cal P}_e}
\newcommand{\polg}{{\cal P}_\gamma}
\newcommand{\poll}{{\cal P}}
\newcommand{\apolel}{\langle{\cal P}_e\rangle}
\newcommand{\ameas}{A_m}
\newcommand{\lum}{{\cal L}}
\newcommand{\alum}{A_{\cal L}}
\newcommand{\aeff}{A_\varepsilon}
\newcommand{\apol}{A_{\cal P}}
\newcommand{\apower}{{\cal A}}
\newcommand{\aengy}{A_E}
\newcommand{\aback}{A_b}
\newcommand{\etal}{{\it et al.}}
\newcommand{\zo}{Z}
\newcommand{\rar}{\rightarrow}

\begin{abstract}
We discuss the ultimate precision for $\ALR$, and therefore for
the weak mixing angle, at a high luminosity Linear Collider.
Drawing on our experience at the SLC, and considering
various machine parameter sets for the NLC and for TESLA, it emerges that
a compromise between peak luminosity and precision will be a likely outcome.
This arises due to the severe requirements on the uncertainty in the luminosity
weighted collision energy ($E_{CM}$).  We consider the cases with and without 
a polarized positron beam.  
\end{abstract}

\section{Introduction}

The determination of the weak mixing angle $\swein$ derived from the $\ALR$ measurement at the SLC
is by far the most precise presently available.  Based on 540 thousand events with a mean electron
beam polarization of $\sim 72\%$ the uncertainty on $\ALR$ consists of an approximately
1.3\% statistical error, and a 0.64\% systematic error (all errors are relative)
leading to a precision of $\pm 0.00027$ on $\swein$\cite{ALRPRL}. The measurement, now statistically
limited, is the only $\swein$ determination that shows promise for sufficiently improved systematic error
to be useful at a future Z factory, where data samples of order $10^9$ events are possible. 

As is well known, the dominant systematic error is due to polarimetry ; 
an uncertainty of $0.50\%$ has been obtained using
a Compton polarimeter.  Less well known is that the next largest systematic error ($0.39\%$)
arises from the conversion of $\ALR(E_{cm})$ to its Z-pole value $\ALR^0$
(and hence to $\swein$)
by correcting for initial state radiation and the contribution and interference
due to the pure photon amplitude.  This calculation requires accurate and precise
knowledge of the luminosity-weighted average center-of-mass collision energy $E_{CM}$.  
A useful rule of thumb is that a 80 MeV error in $E_{CM}$ translates into a
$1\%$ uncertainty in $\ALR^0$.  From this rule, along with the fact that the
statistical error $\delta \ALR = (\poll \sqrt(N))^{-1}$ and $\ALR \sim 0.15$,
it is apparent that a sample of $10^9$ events at an effective polarization of
close to $100\%$ would require an understanding of $E_{CM}$ at the level of 
1 MeV.  We will return to this point following a discussion of all other systematic
effects.

\section{Polarimetry and Polarization}

At a future Linear Collider (LC), two possibilities may be envisioned : either only the electron
beam is polarized (as it was at the SLC), or a suitable positron source can be built
and both beams are polarized.  We will deal with these possibilites in turn.
For the following discussion, we will assume that hoped for improvements
in photocathode technology will provide $90\%$ electron polarization (all results
are easily scalable)\cite{90pol}.

In the event that only the electron beam is polarized, it is likely that a precision
Compton polarimeter would be used\cite{woodspol}.
A Compton polarimeter detects beam electrons
that have been scattered by photons from a circularly polarised laser.
The choice of a Compton scattering polarimeter is dictated by the requirements
that the device be operated continually while beams are in collision
and that uncertainties in the physics of the 
scattering process not be a limiting factor in the systematic error; both 
troublesome issues for M\o ller scattering instruments.  In addition,
the pulse-to-pulse controllability of the laser target polarization (at 120 Hz
at the SLC), as well as the high polarization ($99.9\%$), are additional advantages over 
other options.  Based on our experience with
the SLD Compton polarimeter, how far can this technology be pushed ?
     
\begin{table}
\caption{Systematic uncertainties that affect Compton polarimetry.
The chromaticity and IP corrections are due to the accelerator rather than the polarimeter.}
\label{table1}
\begin{tabular}{lcc}
Uncertainty & SLC $\delta\pole/\pole$~($\%$) & Future LC $\delta\pole/\pole$~($\%$) \\
\hline
Analyzing power calibration & 0.40 & 0.20 \\
Detector linearity & 0.20 & 0.10  \\
Laser polarization & 0.10 & 0.10  \\
Electronic noise & 0.20 & 0.05  \\ 
\hline
Total polarimeter uncertainty  & 0.50 & 0.25  \\
{\it Chromaticity and IP corrections} & 0.15 & {\it negl.} \\
\end{tabular}
\end{table}

Table~\ref{table1} gives a breakdown of instrumental effects, and best estimates for
plausible improvements.  It is assummed that a multichannel electron
spectrometer supplemented by Compton gamma detectors for cross calibration
is used, and that the Compton scattering IP is located downstream of the
$\ee$ collision IP to allow for tests of collisional effects\cite{downstream}. The small
chromaticity effect observed at the SLC is expected to be very small
in a true LC (ie, without arcs), as is any collisional depolarization\cite{depol}.
We conclude that a factor of two improvement over the SLC results is achievable.
Were this the limiting systematic (a likely situation), it would be
possible to improve on the SLD result by a factor of 5, a precision on $\swein$
of about $\pm 0.00005$.  A relatively ``modest'' data sample of 50 million events would be
sufficient in this case.

If positron polarization is available, for arguments sake at the level of $50\%$,
dramatic improvements become possible.  By virtue of the fact that the effective
polarization is very close to $100\%$
 
\begin{equation}
\poll_{eff} = {{\pole + \polp}\over{1 + \pole\polp}} = 96.6\%,
\end{equation}
the fractional uncertainty on $\poll_{eff}$ is small - $0.10\%$ if a
compton polarimeter with $0.25\%$ precision is used for both the
electron and positron beams.  For this very small error, issues of
non-polarimetric systematics become a serious issue
as will be discussed later, but in principle
$\swein$ uncertainties approaching $\pm 0.00002$ are obtainable with a
sample of 100 million Zs.   

By using the ``Blondel'' scheme, whereby all four $\ee$
helicity configurations (LL,LR,RL,RR) are collected\cite{blondel},  
the need for any absolute polarimetry is in principle eliminated.
Typically, only $10\%$ of the collected luminosity needs to
be taken in the low cross section (LL or RR) configuration.
In this technique, luminosity-weighted beam polarizations are 
obtained directly (so that for example, any collisional effects
are accounted for).  Polarimeters will still be needed 
in order to carefully monitor left-right polarization {\it differences}
for each beam at the level of $10^{-3}$.  This capability was demonstrated
at the SLC, where it was advantageous that the electron helicity was 
changed randomly pulse-to-pulse.  It is not clear that helicity
reversals can be performed rapidly enough at
a polarized positron source. Even were they done every
few minutes, larger L/R beam asymmetries and their associated 
uncertainties may occur.  This issue
warrants more detailed study.  Nevertheless,
it may be possible to achieve $\swein$ precision below $\pm 0.00002$
in this way, so long as all non-polarimetric uncertainties can be
held to a total of less than $0.10\%$.  In what follows all
relevant effects are discussed, and we will argue that energy
measurement will pose the greatest challenge.

\section{Other systematics}

The measured asymmetry $\ameas$ is related to $\ALR$ by the following
expression which incorporates a number of small correction terms in
lowest-order approximation, 

\begin{eqnarray}
\label{eq:ALRsys}
\ALR & = & \frac{\ameas}{\apolel}+\frac{1}
{\apolel}\biggl[f_b(\ameas-\aback)-\alum+\ameas^2\apol \nonumber \\
&  & -E_{cm}\frac{\sigma^\prime(E_{cm})}{\sigma(E_{cm})}\aengy
-\aeff + \apolel\polp \biggr], 
\end{eqnarray}
where $\apolel$ is the mean luminosity-weighted polarization; 
$f_b$ is the background fraction;
$\sigma(E)$ is the unpolarized Z boson cross section at energy $E$;
$\sigma^\prime(E)$ is the derivative of the cross section with
respect to $E$;
$\aback$, $\alum$, $\apol$, $\aengy$, and
$\aeff$ are the left-right asymmetries\cite{asymdef}
of the residual background,
the integrated luminosity, the beam polarization,
the center-of-mass energy, and
the product of detector acceptance and efficiency, respectively;
and $\polp$ is any longitudinal positron polarization which is assumed to
have constant helicity\cite{posipol}.

At the SLC backgrounds were understood at the level of $3 \times 10^{-4}$.
While linear colliders are inherently less forgiving than storage rings,
we believe the required performance of $10^{-4}$ or better can be attained.
Luminosity asymmetries ($\alum$) at the SLC were
reduced using feedback at the source and by occasional 
reversals using a spin rotator solenoid,
and were known to approximately $10^{-4}$. 
With improved small angle Bhabha and radiative
Bhabaha detectors, it should be possible to do even better, {\it if the
frequency of helicity reversal for the positrons (discussed above for the case of
$\apol$) does not present difficulties}.  The other asymmetries $\aengy$ (about
$10^{-7}$ at the SLC), and particularly $\aeff$\cite{LReff}, should not present a problem).
We note that in the event positrons are nominally unpolarized, 
the precision expected for this case makes it
necessary to verify this to better than $2 \times 10^{-4}$. At the SLC, a
dedicated experiment achieved a precision of $\delta\polp = 7 \times 10^{-4}$, so this
goal seems reasonable.  

Finally, overlapping Z events may be a complication, in particular in the NLC design.
Even at lower NLC luminosities ($2.6 \times 10^{33}$ at a 120 Hz repetition rate),
there is a $14\%$ probability for 2 or more Z's in a given bunch train.  Studies
are needed to demonstrate that multiple Z events can be easily identified with the
required reliability (for a $>10^8$ event sample, misidentification must be kept below
$0.02\%$ so as not to become a limiting uncertainty).     

\subsection{Energy systematics}

Anyone familiar with the heroic efforts at LEP required to achieve
their spectacular energy uncertainties of order 1 MeV (and a
Z mass error of 2.1 MeV), will appreciate the problem faced at a
future LC.  The method of resonant depolarization will not be an
option at a LC - rather it will be necessary to establish absolute
calibration using precision spectrometers and the Z pole, with additional
relative beam energy data coming from the acolinearity of Bhabha events. 
The required instrumentation, whether it involves SLC-style precision
magnetic spectrometers located in the extraction lines, LEP-style BPM spectrometers,
and possibly wire scanners at positions of high beam dispersion, will be very important
and must be incorporated into any machine design.  
In addition, the machine stability and the (non-gaussian) energy distributions
are less favorable at a LC.

Due to beamstrahlung effects, there is a potentially large difference
between the sum of the undisrupted beam energies (pre-collision), and the luminosity-weighted
$E_{CM}$.  At the SLC, the effect was typically $>40$ MeV during high
luminosity running, and the size of this effect was
very sensitive to machine operating conditions. 
We have calculated beamstrahlung effects for a variety of NLC and TESLA
parameter sets using the program GUINEAPIG\cite{guineapig}.  We have already varified agreement
to about $20\%$ between GUINEAPIG and our energy measurement 
data (using energy spectrometers with $\sim 20$
MeV precision for beams in and out of collision) for the SLC.  In our experience these effects
were relatively unstable in time, and changed significantly as the luminosity was optimized.  
Depending entirely on Z-lineshape scans to incorporate all beamstrahlung effects might be unwise,
as a 1 MeV statistical error on the peak location would require a $\sim 4$ million event 
equivalent integrated luminosity and hence a sizeable fraction of a day even at the highest
envisioned luminosities\cite{zpeak}. It
is therefore a good strategy to minimize beamstrahlung as much as possible, with a reasonable goal of
$10\%$ relative precision on its determination.  

Table~\ref{table2} illustrates the mean energy loss corrections for a representative set of
LC designs.  In the ``nominal'' NLC design
the effect is larger (125 MeV) than
it was at the SLC (49 MeV), while the ``nominal'' TESLA design is more forgiving (44 MeV).
By operating at reduced bunch charge or increased horizontal beta function ($\beta_x$),
beamstrahlung effects can be substantially reduced, albeit at the cost of reduced
luminosity.  We investigated a number of scenarios for NLC and TESLA. For example,
a nine-fold increase in $\beta_x$ will reduce the beamstrahlung energy losses to
18 MeV and 1 MeV respectively, with concurrent losses in luminosity by factors of
4.8 and 4.3 relative to nominal\cite{tor}.  These reduced-luminosity configurations would
probably be required for the highest precision $\ALR$ programs using 1 billion
Z events where $\cal{O}({\rm 2 MeV})$
energy precision is needed.  In general the TESLA design is more favorable for these more
ambitious goals.   

 
\begin{table}
\caption{Some Z-pole machines and parameter sets simulated using GUINEAPIG.
The luminosity-weighted beamstrahlung energy loss corrections are given in the last row.}
\label{table2}
\begin{tabular}{lccccc}
  & SLC & NLC-90 & NLC-90(low) & TESLA-90 & TESLA-90(low) \\
\hline
Luminosity ($10^{33}$ cm$^{-2}$ s$^{-1}$) & 0.0024 & 3.9 & 0.9 & 6.5 & 1.5 \\
Repetition rate (Hz) & 120 & 180 & 180 & 5 & 5  \\
Bunches per train & 1 & 95 & 95 & 2820 & 2820 \\
Bunch charge ($10^{10}$) & 4.0 & 1.0 & 1.0 & 2.0 & 2.0  \\
$\gamma \epsilon_x/\gamma \epsilon_y$ ($10^{-8}$ m-rad) & 6000/1200 & 400/6 & 400/6 & 1000/3 & 1000/3 \\
$\beta_x/\beta_y$ at IP (mm) & 3.6/3.7 & 10/0.10 & 90/0.10 & 15/0.4 & 135/0.4  \\
$\sigma_z$ ($\mu{\rm m}$) & 920 & 125 & 125 & 400 & 400 \\
{\it Lum.Wt. beamstr. E-loss} (MeV) & 49 & 125 & 18 & 44 & 1  \\
\end{tabular}
\end{table}

\section{Conclusions}

A five-fold improvement over the SLD result, to $\delta \swein = \pm 0.00005$, is plausible 
with a ($90\%$) polarized electron beam and about 50 million events.  With both beams polarized ($90\%/50\%$),
an error approaching $\pm 0.00002$ may be possible if energy uncertainties are at the 5 MeV level.
To reach this high precision will probably require a reduced-beamstrahlung/reduced-luminosity
machine configuration, most easily attained in the TESLA design.  In addition, the issue of rapid
reversal of the polarized positron source, necessary for the control of systematic left-right asymmetries,
requires further study.  

We note that the precision electroweak program at a LC would include only modest
improvements in the Z lineshape parameters compared to the LEP results, but might provide a
W mass measurement to better than 10 MeV (6 MeV for $100 \rm{fb}^{-1}$)\cite{monig}.
Many of the energy related issues discussed for
$\ALR$ would apply to a W threshold scan; in addition this measurement would require a
long extrapolation from the Z-pole energy calibration point.  
The ultimate precision electroweak program at a LC would, we think, require somewhat specialized
running conditions and reduced luminosity, either at the Z pole or the W threshold.


\begin{references}
\bibitem{ALRPRL} K.~Abe {\it et al.}  [SLD Collaboration],
Phys.\ Rev.\ Lett.\  {\bf 84}, 5945 (2000).
\bibitem{90pol} A.V. Subashiev and J.E. Clendenin, SLAC-PUB-8312, (2000).
\bibitem{woodspol} M. Woods, SLAC-PUB-8397, (2000).
\bibitem{downstream} A downstream Compton IP is not presently accomodated in
the TESLA design.  For this machine, collisional effects would need to be
controlled using positron polarization and the ``Blondel'' scheme discussed at 
the end of section II.
\bibitem{depol} P.~Chen and K.~Yokoya, {\it Proceedings of the Eighth International 
Symposium on High-Energy Spin Physics}, Minneapolis, MN, 1988, pg.~938.
\bibitem{blondel} A. Blondel, Phys.Lett. B202,145 (1988).
\bibitem{asymdef} The left-right asymmetry for a quantity $Q$ is defined as
$A_Q\equiv(Q_L-Q_R)/(Q_L+Q_R)$ where the subscripts $L$,$R$ refer to
the left- and right-handed beams, respectively.
\bibitem{posipol} Since the colliding electron and positron bunches
were produced on different machine cycles and since the
electron helicity of each cycle was chosen randomly, any positron
helicity arising from the polarization of the production electrons
was uncorrelated with electron helicity at the IP. However,
positron polarization of constant helicity would affect the measurement.
\bibitem{LReff} The value of $\ALR$ is unaffected by decay-mode-dependent
variations in detector acceptance and efficiency
provided that the efficiency for detecting a fermion at
some polar angle (with respect to the electron direction)
is equal to the efficiency for detecting
an antifermion at the same polar angle.  Charge-sign dependent
detection efficiency should be an extremely small effect in any 
operating or planned collider detector.
\bibitem{guineapig} D. Schulte, Ph.D. thesis, Hamburg, 1996.
\bibitem{zpeak} At the SLC, about 10K event equivalents were used in an optimized 
three-point peak scan to achieve a 20 MeV statistical error.
\bibitem{tor} Updated NLC machine parameter sets, somewhat different then those
given here, were presented by Tor Raubenheimer (see these proceedings).
\bibitem{monig} K. M$\ddot {\rm o}$nig, these proceedings. See also R. Hawking and K. M$\ddot {\rm o}$nig,
DESY 99-157, (1999).
\end{references}
\end{document}